\documentclass[epjCONF]{svjour}
\usepackage{graphics}
\usepackage[varg]{txfonts} % Times fonts
\usepackage[latin1]{inputenc}
\session-title{CNR*09}
\begin{document}
\title{Nucleus-nucleus potential, energy dissipation
and mass dispersion in fusion and transfer reactions}
\author{Kouhei Washiyama\inst{1}\fnmsep\thanks{\email{washiyama@ganil.fr}} 
\and Denis Lacroix\inst{1}\fnmsep\thanks{\email{lacroix@ganil.fr}}
\and Sakir Ayik\inst{2}\fnmsep\thanks{\email{ayik@tntech.edu}} }
\institute{GANIL, Bd Henri Becquerel, BP55027, 14076 Caen, France 
\and Physics Department, Tennessee Technological University, Cookeville, Tennessee 38505, USA}
\abstract{
The nucleus-nucleus potential and energy dissipation in fusion reactions
are obtained from microscopic mean-field dynamics.
The deduced potentials nicely reproduce the one extracted from experimental data.
Energy dissipation shows a universal behaviour between different reactions.
Also, the dispersion of mass distribution in transfer reaction
is investigated in a stochastic mean-field dynamics.
By including initial fluctuations in collective space, 
the description of the dispersion is much improved compared to that of mean field only. 
The result is consistent with the macroscopic 
phenomenological analysis of the experimental data.
} 
\maketitle
\section{Introduction}
\label{intro}

The interplay between nuclear structure and dynamical effects is crucial to properly 
describing nuclear reactions at energies close to the Coulomb barrier.
Therefore, the theories describing such nuclear reactions need 
a unified description for both nuclear structure and dynamics.
Moreover, recent developments on nuclear facilities
introduce much interest on the properties of nuclei far from the stability.
The time-dependent Hartree-Fock (TDHF) theory \cite{bonche76,koonin80,negele82,Sim08}
based on the Skyrme energy density functional (EDF) provides a rather 
unique tool for describing nuclei over the whole nuclear chart.
The TDHF theory solves the time evolution of single-particle wave functions
according to 
\begin{eqnarray}
\label{eq:tdhf}
i\hbar \frac{\partial }{\partial t}\rho = [h[\rho],\rho],
\end{eqnarray}
where $h[\rho]$ denotes the self-consistent mean-field Hamiltonian from the Skyrme EDF, 
denoted by $\cal E[\rho]$, obtained from $h[\rho]=\delta {\cal E[\rho]}/\delta \rho$
with the one-body density $\rho$.
This model automatically includes important dynamical effects such as vibrations of nuclei,
neck formations, and nucleon transfer during reactions.
Since recent computational developments now enable us to 
include all the terms of the Skyrme EDF used in static Hartree-Fock calculations 
in the three-dimensional coordinate space \cite{kim97,nakatsukasa05,umar06b,maruhn06}, 
the description of nuclear reactions using TDHF should be revisited.

In this contribution, 
as an illustration of applications of the TDHF theory to nuclear reactions,
we discuss the properties of nucleus-nucleus potential and 
energy dissipation extracted from the TDHF model \cite{washiyama08,washiyama09a}. 
Moreover, we investigate fluctuations of one-body observables, especially,
the dispersion of mass distributions in transfer reactions
using a stochastic mean-field model \cite{ayik09,washiyama09b}.

\section{Nucleus-nucleus potential and energy dissipation from mean-field dynamics}
\label{sec:potential}

Nucleus-nucleus potential and energy dissipation
are extracted as follows~\cite{washiyama08,washiyama09a}:
(i)~The TDHF equation for head-on collision is solved 
to obtain the time evolution of the total density of colliding nuclei.
(ii)~From the total density, we compute at each time the  
relative distance $R$, associated momentum $P$, and reduced mass $\mu$ of colliding nuclei.
(iii)~We assume that the time evolutions of $R$ and $P$ obey a classical equation of motion
including a friction term which depends linearly on the velocity $\dot{R}$:
\begin{eqnarray}
\frac{dR}{dt}=\frac{P}{\mu },~~~~
\frac{dP}{dt}=-\frac{dV}{dR}-\gamma (R)\dot{R},
\label{eq:newtonequation}
\end{eqnarray}
where $V(R)$ and $\gamma (R)$ denote the nucleus-nucleus potential 
and friction coefficient, respectively. 
The friction coefficient $\gamma (R)$ describes the effect of energy dissipation 
from the macroscopic collective degrees of freedom to the microscopic ones.
For the TDHF calculations presented in this contribution, 
the three-dimensional TDHF code developed by P.~Bonche 
and coworkers with the SLy4d Skyrme effective force~\cite{kim97} is used.
The mesh sizes in space and in time are 0.8~fm and 0.45~fm/$c$, respectively.
For more details, see Refs.~\cite{washiyama08,washiyama09a}.

\begin{figure}[tbhp]
\begin{center}\leavevmode
\resizebox{0.9\columnwidth}{!}{%
  \includegraphics{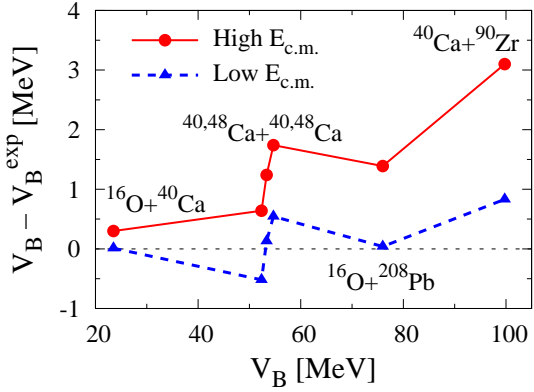} }
\caption{
Barrier height $V_B$ deduced from our method minus experimental barrier 
height $V_B^{\rm exp}$ as a function of extracted barrier height 
for the reactions indicated in the figure. 
The values $V_B$ are deduced from high energy (solid line) and from low energy (dashed line)
TDHF trajectories, respectively.
}
\label{fig:systematics}
\end{center}
\end{figure}

Dynamical effect on potentials deduced from TDHF trajectories at center-of-mass energies
close to the Coulomb barrier is seen in all reactions considered here. 
Figure~\ref{fig:systematics} shows the difference between the barrier height 
deduced from TDHF trajectories ($V_B$) and 
the barrier height extracted from experimental data ($V_B^{\rm exp}$) \cite{vaz81,newton04} 
as a function of $V_B$. 
The solid line corresponds to the barrier height extracted using 
high-energy TDHF trajectories ($E_{\rm c.m.}\gg V_B$),
whereas the dashed line is the result for low-energy TDHF trajectories ($E_{\rm c.m.}\sim V_B$).
The former identifies with the barrier height of the frozen density apploximation \cite{denisov02}. 
Dynamical reduction of the barrier height from high-energy TDHF to low-energy TDHF
is clearly seen for all reactions. 
Moreover, because of this reduction, 
the value of the barrier height at low energy approaches the experimental data. 
This underlines the importance of dynamical effects close to the Coulomb barrier
and shows the precision of our method.

\begin{figure}[tbhp]
\begin{center}\leavevmode
\resizebox{0.9\columnwidth}{!}{%
  \includegraphics{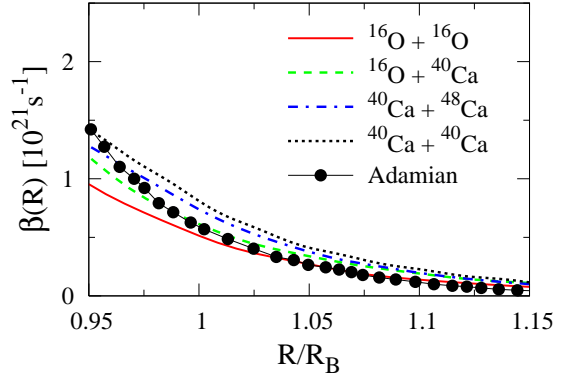} }
\caption{
Extracted reduced friction $\beta(R)\equiv\gamma(R)/\mu $ as a function of $R$ scaled by
the Coulomb barrier radius $R_B$ for different reactions. A microscopic friction coefficient 
by Adamian {\it et al.} is shown by the solid circles for comparison.
}
\label{fig:friction}
\end{center}
\end{figure}

Our method is also able to provide information on energy dissipation
through the friction coefficient $\gamma$, which is shown in Fig.~\ref{fig:friction}.
In this figure, we present reduced friction coefficients 
$\beta (R)\equiv\gamma (R)/\mu $ as a function of $R$ 
scaled by the Coulomb barrier radius $R_B$ for different reactions.
Figure~\ref{fig:friction} clearly shows that the order of magnitude of $\beta(R)$
and the radial dependence are almost independent on the size and asymmetry of the system.
We also compare our results with that of a microscopic model based on small amplitude response
by Adamian {\it et al.} \cite{adamian97} by the solid circles. 
The radial dependence and the magnitude of the friction coefficient are very similar to
those extracted from our method.

From here, we conclude that the use of the macroscopic equation~(\ref{eq:newtonequation})
is valid and mean-field dynamics gives good descriptions 
for the nucleus-nucleus potential and energy dissipation.

\section{Mean-field fluctuations}

It is well known that the mean-field model gives good descriptions for average
evolution of one-body observables in low-energy nuclear reactions. However,
it completely fails in the description of the dynamics of fluctuations of one-body observables.
One of the shortcomings may be the failure of the mean-field description of the dispersion 
of mass distributions of final fragments in deep inelastic collisions.
It has been recognized for a long time that 
TDHF calculations severely underestimate the dispersion
of mass distributions of experiments \cite{davies78,dasso79},
although TDHF calculations well reproduce the mean value of fragment mass.

During the past decades, much effort has been devoted to overcoming this difficulty and 
to developing transport theories that are able to describe not only mean values 
but also fluctuations (for a review, see Refs~\cite{Abe96,Lac04}). 
Among them, the variational principle by Balian and V\'en\'eroni (BV)
appears as one of the most promising methods \cite{Bal84,balian84,Mar85}. 
However, even nowadays it remains difficult to apply \cite{broomfield08}. 
More than 30 years after the first application of the TDHF theory, 
the absence of a practical solution to include fluctuations beyond mean field in a fully microscopic 
framework strongly restricts applications of mean-field-based theories. 

In order to overcome this difficulty, recently
we proposed a stochastic mean-field (SMF) approach, 
which is a stochastic extension of the mean-field model for low energy nuclear dynamics 
so as to include zero-point fluctuations of the initial state \cite{ayik09,ayik08}.
The initial density fluctuations are simulated by representing the initial state in terms of 
a suitable ensemble of initial single-particle density matrices,
which is similar to the idea in Refs.~\cite{esbensen78,dasso92}.
In fact, this idea can be regarded as the beginning of 
constructing time-dependent version of configuration mixing calculations.
In this manner, the description with single Slater determinant is replaced by
a superposition of multi Slater determinants.
A member of the ensemble, indicated by event label $\lambda$, can be expressed as
\begin{eqnarray}
\label{eq:density}
\rho^\lambda({\bf r},{{\bf r}}^{\prime},t) = \sum\limits_{ij \sigma\tau}
\Phi_{i \sigma\tau}^\ast({\bf r},t;\lambda )\rho_{ij}^\lambda(\sigma\tau)
\Phi_{j \sigma\tau}({\bf r}^{\prime},t;\lambda ),
\end{eqnarray}
where summations $i$ and $j$ run over a complete set of single-particle wave functions
$\Phi_{i \sigma\tau}({\bf r},t;\lambda )$, and  $\sigma$ and $\tau$ denote
spin and isospin quantum numbers. According to the description of the SMF approach \cite{ayik08},
the elements of density matrices $\rho_{ij}^\lambda(\sigma\tau)$
are assumed to be time-independent random Gaussian numbers with mean value
$\overline{\rho_{ij}^\lambda(\sigma\tau)}=\delta_{ij}n_i^{\sigma\tau}$
and with the variance of the fluctuating part $\delta\rho_{ij}^\lambda(\sigma\tau)$
specified by
\begin{eqnarray}
&&\overline{\delta\rho_{ij}^\lambda(\sigma\tau)
\delta\rho_{j'i'}^\lambda( {\sigma}'{\tau}')}\nonumber \\
&=&\frac{1}{2}\delta_{j{j}'}\delta_{i{i}'}\delta_{\tau {\tau}'}\delta_{\sigma {\sigma}'}
\left[n_i^{\sigma\tau}(1 - n_j^{\sigma\tau}) + n_j^{\sigma\tau}(1 - n_i^{\sigma\tau})\right].
\label{variance}
\end{eqnarray}
Here, $n_i^{\sigma\tau}$ denotes the average single-particle occupation factor. At zero temperature
occupation factors are $0$ and $1$, and at finite temperature they are determined 
by the Fermi-Dirac distribution.
The great advantage of the SMF approach is that each Slater determinant
evolves independently from each other following the time evolution
of its single-particle wave functions in its self-consistent mean-field Hamiltonian, 
denoted by $h(\rho^\lambda)$, according to
\begin{eqnarray}
\label{eq:spwf}
i\hbar \frac{\partial }{\partial t}\Phi_{i \sigma\tau} ({\bf r},t;\lambda )
= h(\rho^\lambda )\Phi_{i \sigma\tau}({\bf r},t;\lambda ).
\end{eqnarray}

In the following applications, we focus on the head-on $^{40}$Ca+$^{40}$Ca collision
around the Coulomb barrier energy.

\subsection{Fusion reactions}

\begin{figure}[btph]
\begin{center}\leavevmode
\resizebox{0.85\columnwidth}{!}{%
  \includegraphics{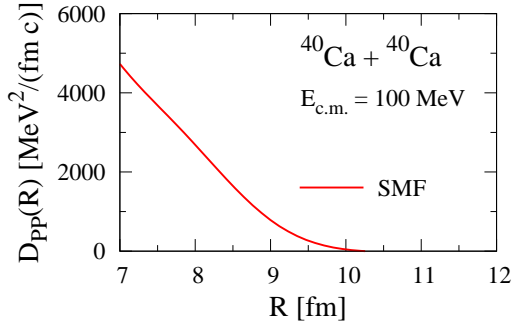} }
\caption{
Diffusion coefficient as a function of the relative distance
for the head-on $^{40}$Ca+$^{40}$Ca collision at $E_{\rm c.m.}=100$~MeV.
}
\label{fig:Dpp}
\end{center}
\end{figure}

First, we apply the SMF approach to fusion reactions \cite{ayik09}. 
To discuss the fluctuation of collective variables,
we map the SMF time evolution to a one-dimensional macroscopic Langevin equation,
which is similar to Eq.~(\ref{eq:newtonequation}) except
an additional Gaussian random force $\xi_P^\lambda(t)$: 
\begin{eqnarray}
\label{eq:langevin} \frac{d}{dt}P^{\lambda} = - \frac{d}{dR^{\lambda}}U(R^{\lambda} ) 
- \gamma (R^{\lambda} )\dot {R}^{\lambda} + \xi_P^{\lambda} (t),
\end{eqnarray}
Ignoring non-Markovian effects, the random force $\xi_P^\lambda(t)$ with zero mean value 
reduces to white noise specified by a correlation function,
\begin{eqnarray}
\label{eq:correlation}
\overline{\xi _P^\lambda (t)\xi _P^\lambda({t}')} = 2\delta(t-{t}')D_{PP}(R).
\end{eqnarray}
Here $D_{PP}(R)$ denotes the momentum diffusion coefficient.
We note that the expression of the diffusion coefficient has the same form
as that obtained from the phenomenological nucleon exchange model ~\cite{feldmeier87}.
As an example, the diffusion coefficient for the head-on $^{40}$Ca+$^{40}$Ca
collision at $E_{\rm c.m.}=100$~MeV is shown in Fig.~\ref{fig:Dpp}.

\subsection{Mass dispersion in transfer reactions}

Another application has been made to the dispersion of the fragment mass distribution
to improve the mean-field description~\cite{washiyama09b}.
To do so, we investigate transfer reactions near the Coulomb barrier,
where nucleon exchange will occur during reaction,
and estimate the dispersion of the fragment mass distribution.

In the SMF approach, time evolution of the mass number of the projectile-like fragment $A_P^\lambda $ 
is also described by a Langevin equation \cite{Randrup2},
\begin{eqnarray}
\label{eq:langevin-mass}
\frac{d}{dt}A_P^\lambda = v(A_P^\lambda, t) + \xi_A^\lambda(t),
\end{eqnarray}
where $v(A_P^\lambda ,t)$ denotes the drift term for nucleon transfer.
The Gaussian white noise random force $\xi_A^\lambda (t)$ is determined with zero mean value 
and with a correlation function,
\begin{equation}
\overline{\xi_A^\lambda (t)\xi_A^\lambda ({t}')}= 2\delta(t -{t}')D_{AA},
\end{equation}
where $D_{AA}$ is the diffusion coefficient associated with nucleon exchange.
The variance $\sigma_{AA}^{2}$ of fragment mass distribution is determined 
by small fluctuations of the mass number $\delta A_{P}^{\lambda}$ through 
$\sigma_{AA}^{2}(t)=\overline{\delta A_{P}^{\lambda}\delta A_{P}^{\lambda}}$. 
According to the Langevin equation, neglecting
contributions from the drift term, the variance is related to the diffusion
coefficient according to \cite{Randrup2,Randrup82}
\begin{eqnarray}
\sigma^2_{AA}(t) \simeq 2 \int_0^t D_{AA}(s)ds.
\label{eq:sigma}
\end{eqnarray}

In the phenomenological nucleon exchange model, 
the relation $ \sigma_{AA}^{2}(t)=N_{\rm exc}(t)$ was obtained,
where $N_{\rm exc}(t)$ denotes the accumulated total number of exchanged nucleons until time $t$,
and was extensively used to analyze the experimental data of mass dispersion \cite{Fre84}.
In the following, to check whether the SMF approach satisfies the above relation,
we estimate the both quantities by the SMF approach.

\begin{figure}[bthp]
\begin{center}\leavevmode
\resizebox{0.95\columnwidth}{!}{%
  \includegraphics{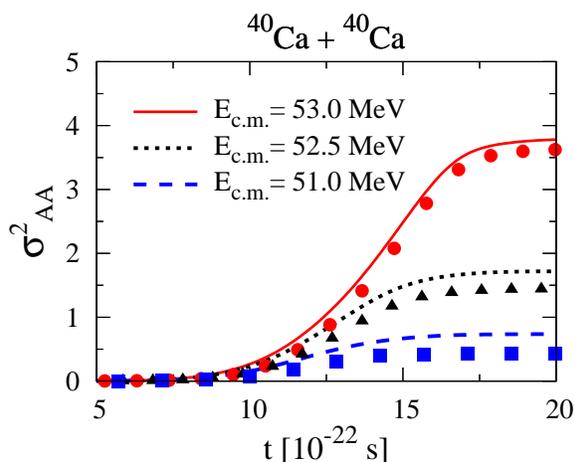} }
\caption{
Time evolution of $\sigma_{AA}^2$ obtained from the SMF approach for $^{40}$Ca+$^{40}$Ca
reaction at different center-of-mass energies.
Number of exchanged nucleon is superimposed by the solid circles, solid triangles, and
solid squares from high to low energies.
}
\label{fig:SIGMAtime}
\end{center}
\end{figure}

In Fig.~\ref{fig:SIGMAtime},
the variances of the fragment mass distributions deduced from the SMF approach 
for the head-on $^{40}$Ca+$^{40}$Ca reaction at three center-of-mass energies
are shown by lines.
The number of exchanged nucleons is inserted in Fig.~\ref{fig:SIGMAtime} by 
the solid circles, solid triangles, and solid squares from high to low energies.
The mass variance estimated from the SMF approach is consistent 
with this relation.
We also estimate the variance of fragment mass distribution using the standard TDHF approach. 
The asymptotic values of $\sigma_{AA}^2$ for the $^{40}$Ca+$^{40}$Ca reaction 
are 0.004, 0.008, and 0.008 from low to high energies, 
while the number of exchanged nucleons are 0.432, 1.441, and 3.634.
The TDHF results are much smaller than the number of exchanged nucleons and
are also much smaller than the results obtained from the SMF approach 
that are 0.730, 1.718, 3.790.
The failure of the TDHF theory on the description of variances of the fragment mass distribution
has been recognized for a long time as a major limitation of the mean-field theory. 
It appears that the SMF approach cures this shortcoming.
As seen from Fig.~\ref{fig:SIGMAtime}, 
not only the asymptotic value of $\sigma^2_{AA}$ but also the entire
time evolution is very close to the evolution of $N_{\rm exc}(t)$. 

\section{Conclusion}

Mean-field dynamics and mean-field fluctuations using microscopic time-dependent models
are discussed in the context of low energy nuclear reactions.
We have shown that the TDHF theory gives precise values of nucleus-nucleus
potential and a universal behavior of energy dissipation.
We have also shown that the SMF approach correctly
describes the mass dispersion of final fragments in transfer reaction
at energies near the Coulomb barrier.
This gives a practical solution to properly describe mean-field fluctuations
on top of mean field.

\begin{acknowledgement}
 This work is supported in part by US DOE Grant DE-FG05-89ER40530.
\end{acknowledgement}

\end{document}